\documentclass[11pt]{article}

\usepackage[margin=1in]{geometry}
\usepackage[T1]{fontenc}
\usepackage{textcomp}
\usepackage{graphicx}
\usepackage{booktabs}
\usepackage{multicol,multirow}
\usepackage{amsmath,amssymb,amsfonts}
\usepackage{mathrsfs}
\usepackage{amsthm}
\usepackage{newtxtext}
\usepackage{algorithm}
\usepackage{algpseudocode}
\usepackage{rotating}
\usepackage{appendix}
\usepackage{xcolor}
\usepackage{setspace}
\usepackage[backend=biber,style=apa,natbib=true,sortcites=true]{biblatex}
\usepackage[
  colorlinks,
  allcolors=blue,
  pdfauthor={Wenqian Xu and Feng Ji},
  pdftitle={Reinforcement Learning Measurement Model},
  pdfsubject={},
  pdfkeywords={Reinforcement Learning, Process Data, Markov Decision Process, Measurement Model, Scalability, Item Response Theory}
]{hyperref}

\DeclareLanguageMapping{english}{english-apa}

\theoremstyle{definition}

\numberwithin{equation}{section}

\newcommand{\cS}[1]{#1}

\title{Reinforcement Learning Measurement Model}
\author{%
  \begin{tabular}{cc}
    \textbf{Wenqian Xu}\textsuperscript{1} &
    \textbf{Feng Ji}\textsuperscript{1}
  \end{tabular}\\[0.7em]
  {\normalsize\textsuperscript{1}Department of Applied Psychology and Human Development,
  University of Toronto, ON, Canada}\\[0.45em]
  {\normalsize\textbf{Corresponding author:} Feng Ji \texttt{<f.ji@utoronto.ca>}}
}
\date{}

\begin{document}

\maketitle

\begin{abstract}
Interactive assessments generate sequential process data that are not well handled by conventional item response models. Existing MDP-based measurement approaches, such as the Markov decision process measurement model \citep[MDP-MM,][]{lamar2018markov}, link action choices to state-action values, but their reliance on person-specific tabular value functions makes them difficult to scale beyond small, fully enumerated tasks. We propose the Reinforcement Learning Measurement Model (RLMM), a measurement framework that decouples person-level choice sensitivity from task-level value representation through a shared parametric action-value function, making estimation more computationally efficient for larger process-data settings. The model combines a Boltzmann choice rule with normalized advantages, a soft Bellman consistency penalty, and a block-coordinate MAP procedure for joint estimation, while also yielding step-level influence diagnostics for identifying behaviorally critical decisions. In peg-solitaire simulations, the RLMM achieved higher estimation accuracy and substantially lower runtime than the original MDP-MM, with advantages increasing as task complexity grew. In AQUALAB gameplay logs, the estimated person parameter was positively associated with cumulative reward, task completion, and behavioral efficiency. These results show that the RLMM extends decision-process-based psychometric models to larger and more behaviorally realistic environments while preserving an interpretable latent trait tied to decision making steps.
\end{abstract}


\section{Introduction} \label{sec:intro}

Item response theory (IRT) provides a popular statistical framework for measuring proficiency from assessment data by linking observed item responses to person parameters and item characteristics such as difficulty and discrimination \citep{rasch1960probabilistic,birnbaum1968latenttrait,lord1980applications}. In its classical form, IRT treats each item as a measurement opportunity and is well matched to tests in which each item yields a single final response. The framework was later extended beyond dichotomous scoring to accommodate richer response formats; for example, the nominal response model represents an item response as a multinomial choice among categories \citep{bock1972nominal}. However, these developments still take the realized response to an item as the basic unit of analysis. This abstraction becomes limited in interactive environments, where performance unfolds through a sequence of state-dependent decisions rather than a single terminal answer.

Technology-based assessments increasingly rely on interactive problem-solving tasks in which examinees interact with a dynamic environment. As a result, assessment platforms now record detailed process data, such as time-stamped action logs, state transitions, and intermediate outcomes. These data can support more informative measurement than models based only on final correctness or total scores. While conventional IRT benefited from estimation frameworks such as marginal maximum likelihood that made large-scale calibration operationally feasible \citep{bockaitkin1981mml}, it is less clear how to extend the same combination of interpretability and scalability from item-level responses to action-sequence data. We therefore seek a measurement model that (i) extracts stable latent traits (e.g., ability, and potentially strategy or speed) from action-sequence data, (ii) preserves interpretability at the level of tasks and steps, and (iii) remains computationally feasible as the state space and behavioral heterogeneity grow.

One common approach in this regard is to recode process data as auxiliary indicators and analyze those summaries with standard measurement models, rather than model the state-dependent sequence itself. In cognitive diagnosis modelling literature, early studies used response times to improve attribute classification or to measure fluency jointly with accuracy \citep{Zhan2018RTCDM,WangChen2020Fluency}. Later work added fixation counts as another source of diagnostic information \citep{Zhan2022RTFixationCDM}. \citet{Liang2023ProcessDataCDM} follow the same logic by combining response times and mouse click and drag traces with a conventional cognitive diagnosis model. Item expansion applies this idea at the task level: a single interactive task is recoded as several process-based indicators and then analyzed with a diagnostic classification model \citep{Zhan2022ItemExpansion}. Related work also treats strategy as an explicit latent variable in the cognitive diagnosis model, allowing for multiple strategies, shifts in strategy over time, or response-time-informed strategy selection \citep{Zhang2021MMSDINA,Liao2023StrategyShift,Wei2024MSCDMRT}. These models remain close to standard psychometric practice, but they still do not represent the path an examinee follows through the task state by state.

A different direction is to model process data sequentially, using a generative model for state-action sequences. Markov-IRT approaches use IRT-type models to capture local sequential structure in problem-solving processes \citep{Shu2017MarkovIRT}. Sequential response models make the task's state graph explicit: logs are mapped to discrete states, allowable transitions are enumerated, and transition-level parameters yield interpretable diagnostics about where examinees deviate from optimal or goal-directed pathways \citep{Han2022SRM}. Recent extensions bring covariates into the framework to study group differences and measurement fairness \citep{Han2026SRMCovariates}, and add growth parameters to capture how response tendencies change with accumulated experience during the task \citep{Han2026SRMGrowth}. Other extensions broaden the measurement target and the data modality, including multidimensional sequential response models that infer multiple correlated abilities from state sequences \citep{Han2025MSRM}, joint hierarchical models that simultaneously capture action choices and action times \citep{Fu2024JointSeqTime}, and dynamic choice formulations that jointly model subsequent actions and their timing within continuous-time measurement frameworks \citep{Chen2020CTDynamicChoice}. Complementary probabilistic approaches compress event streams into latent behavioral patterns evolving under Markov dynamics \citep{Xu2020LatentTopicMarkov}, and latent Markov models that relax measurement invariance to avoid confounding heterogeneity across items and events with latent-state change \citep{Kang2025NoninvariantLMM}. Taken together, these models shift the unit of analysis from items to steps, transitions, and time, and they produce parameters that can be interpreted at the level of within-task behavior.

Despite this progress, it remains unclear how known task dynamics can be incorporated into sequential measurement models in a way that improves interpretability and scalability. Many existing models for process data extend item-response logic to action sequences by discretizing states, actions, and allowable transitions, and then estimating transition utilities or difficulty parameters from observed trajectories. This approach captures sequential dependence, but it does not necessarily exploit situations in which the task designer already knows the state-action structure and the goal of the task. In such settings, the central measurement question is not only whether an examinee produces a particular action sequence, but whether their actions tend to select moves that are better according to the task dynamics. This motivates a model in which known transition rules are used to define action values, and person parameters are tied directly to the tendency to choose higher-value actions. This view is closely related to work in cognitive science that treats behavior as the outcome of possibly noisy planning in structured Markov decision process (MDP) or partially observable Markov decision process (POMDP) environments, and models inference as recovering latent internal variables from action trajectories. For example, inverse-planning approaches emphasize that beliefs about how the environment works and preferences over outcomes are intertwined and may need to be inferred jointly for stable interpretation \citep{baker2011btom}. In educational contexts, similar ideas underlie MDP-based measurement models, which interpret learner actions through a softmax policy and infer latent knowledge as beliefs over task dynamics rather than as item-level difficulty parameters \citep{rafferty2015inferring}.

A principled bridge between MDP and psychometric measurement is the Markov decision process measurement model \citep[MDP-MM,][]{lamar2018markov}. For example, \citet{ramanarayanan-lamar-2018-toward} formulate dialog interactions as an MDP and use a Boltzmann policy with inverse-temperature as a conversational ability parameter. The MDP-MM treats the task's transition and reward components as fixed and interprets the inverse-temperature parameter in a Boltzmann (softmax) policy as an examinee ability trait: higher ability corresponds to a stronger tendency to select actions with higher state-action values. However, the original MDP-MM computes a separate tabular value function $Q(s,a \mid \beta)$ for each candidate ability value, so the person parameter is coupled with the value function rather than applied to a shared task-level scoring function. During estimation, the likelihood must be evaluated over many candidate ability values, and each candidate requires solving a new dynamic-programming problem to obtain its own table of state-action values. As the state space grows, this repeated computation becomes expensive. Because action values are stored in separate ability-specific tables, the model also does not naturally support a shared parametric or function-approximation form of $Q(s,a)$ that generalizes across states. This makes the original formulation difficult to scale to larger assessment environments, where tabular enumeration is often impractical.

To address these limitations, we propose the reinforcement learning measurement model (RLMM), a scalable framework that preserves the MDP-MM's core logic of using latent ability to model value-based choices while separating the shared task-value representation from person-level choice sensitivity. Instead of relying on tabular value functions $Q(s,a \mid \beta)$, our method uses a shared parametric action-value function $Q_\theta(s,a)$, works with centered and globally normalized advantages to identify the shared score scale, and lets $\beta_j$ govern the strength of a person's preference for higher-valued actions. We learn the shared value function directly from observed trajectories and enforce known task dynamics using a soft Bellman-consistency penalty. This approach yields an interpretable model in which person parameters continue to quantify the propensity to choose higher-valued actions, while function approximation allows the model to scale to larger state spaces. In addition to parameter estimation, the framework offers interpretable step-level diagnostics. By measuring which states and transitions most strongly influence ability inference, it supports the identification of critical task steps. We evaluate the approach in two settings: peg-solitaire simulations, which provide controlled puzzle tasks with known ability parameters and increasingly large state spaces, and AQUALAB gameplay logs, a real-world educational science-game dataset with open-ended student actions. The simulations show improved person-parameter recovery and lower runtime than the original MDP-MM, with the computational gap widening as task complexity increases. The same approach also remains workable on larger environments for which the tabular baseline is no longer practical. In the empirical study, the estimated person parameter is positively associated with reward accumulation, task completion, and efficiency. By decoupling the MDP-MM's measurement logic from person-specific tabular value functions, RLMM retains its psychometric interpretability while extending its applicability to complex, high-dimensional environments.

The rest of the paper proceeds as follows. Section~\ref{sec:background} reviews the relevant reinforcement learning and MDP background and revisits the MDP-MM. Section~\ref{sec:method} introduces RLMM together with its estimation procedure and diagnostic tools. Sections~\ref{sec:experiments} and~\ref{sec:empirical} present the simulation and empirical studies, and Section~\ref{sec:discussion} closes with discussion, limitations, and future directions.

\section{Background} \label{sec:background}

\subsection{Reinforcement Learning and Markov Decision Processes}

A Markov decision process (MDP) is a standard model for sequential decision making under uncertainty. We write an MDP as $(S,A,T,R,\gamma)$, where $S$ is the state space, $A$ the set of actions, $T(s' \mid s,a)$ the transition kernel from state $s$ to $s'$ given action $a$, $R(s,a,s')$ the immediate reward, and $\gamma$ a discount factor. The associated state-action value function (or $Q$-function) assigns to each state-action pair the expected discounted return,
\begin{equation}
Q(s,a)
= \mathbb{E}\!\left[ \sum_{t=0}^{\infty} \gamma^{t} r_t \,\middle|\, s_0 = s,\, a_0 = a \right].
\label{eq:qfunction}
\end{equation}

Within this framework, an agent selects actions according to a policy that maps states to action probabilities.
 In models of human behavior, policies are typically used to represent decision making in sequential tasks, where each choice both depends on the current state and influences future states.
 Because human behavior is typically noisy rather than strictly optimal, action selection is often modeled using a stochastic decision rule, most commonly the Boltzmann (softmax) policy:
\begin{equation}
p(a \mid s) \propto \exp\!\left\{ \beta\, Q(s,a) \right\},
\label{eq:softmax}
\end{equation}
where the inverse-temperature parameter $\beta \ge 0$ quantifies the extent to which higher-valued actions are preferred over lower-valued ones. Larger values of $\beta$ correspond to more consistent, value-aligned choices, whereas values near zero approximate random responding. In this formulation, $\beta$ provides a scalar index of an individual's decision consistency.

\subsection{The Markov Decision Process Measurement Model}

The softmax rule in Eq.~\eqref{eq:softmax} can be reinterpreted as a latent-trait measurement model by treating the inverse-temperature parameter as a person parameter and the action values as task-defined scores. \citet{lamar2018markov} develops this idea into the Markov Decision Process Measurement Model (MDP-MM), in which the goal, transition, and reward components of the MDP are fixed at the task level, and individual differences are represented by a person-specific ability parameter $\beta_j$. Given the task model, the value function $Q(s,a \mid \beta_j)$ is computed for each $\beta_j$ using dynamic programming, and the probability that student $j$ selects action $a$ in state $s$ is
\begin{equation}
p(a \mid s, \beta_j)
=
\frac{\exp\!\left\{ \beta_j\, Q(s,a \mid \beta_j) \right\}}
{\sum_{a' \in A_s} \exp\!\left\{ \beta_j\, Q(s,a' \mid \beta_j) \right\}},
\label{eq:mdpmm}
\end{equation}
where $\beta_j$ is interpreted as the student’s ability to choose higher-valued actions. Structurally, Eq.~\eqref{eq:mdpmm} parallels the nominal response model: for a given state $s$, the action values $Q(s,a \mid \beta_j)$ act as scoring parameters for the available response options, and $\beta_j$ plays the role of a latent trait governing the tendency to select higher-valued actions.

A distinctive feature of the MDP-MM is the person-specific value function: for each $\beta_j$, the Q-function $Q(s,a \mid \beta_j)$ is obtained by solving the MDP under the softmax policy with parameter $\beta_j$. Consequently, the ability parameter enters the model both through the value function $Q(s,a \mid \beta_j)$ and through the choice probabilities in Eq.~\eqref{eq:mdpmm}.

Estimation in the MDP-MM proceeds by marginal maximum likelihood. Let $\mathbf{y}_j$ denote the observed sequence of state--action pairs for student $j$, and assume that the ability parameter follows a log-normal distribution $\beta_j \sim \log N(\mu,\sigma^2)$ with density $g(\beta;\mu,\sigma^2)$. Writing $R$ for the vector of reward parameters that define the task-specific value function, the marginal likelihood for $(\mu,\sigma^2,R)$ is
\begin{equation}
L(\mu,\sigma^2,R)
=
\prod_{j}
\int
p(\mathbf{y}_j \mid \beta;R)\,
g(\beta;\mu,\sigma^2)\, d\beta,
\label{eq:mdpmm_marglik}
\end{equation}
where $p(\mathbf{y}_j \mid \beta;R)$ is obtained from Eq.~\eqref{eq:mdpmm} using the value function $Q(s,a \mid \beta)$ implied by the MDP.

The integral in Eq.~\eqref{eq:mdpmm_marglik} does not admit a closed-form solution because the likelihood depends on $Q(s,a \mid \beta)$, which must be computed iteratively for each value of $\beta$. \citet{lamar2018markov} approximated the integral by Gaussian quadrature over $\lambda_j = \log \beta_j$ using a finite set of quadrature nodes. For each candidate parameter vector $(\mu,\sigma^2,R)$ and each quadrature node, the MDP value function was recomputed via policy iteration until convergence. The resulting numerical approximation to $L(\mu,\sigma^2,R)$ was then maximized over $(\mu,\sigma^2,R)$, analogous to marginal maximum likelihood estimation in item response theory, but with dynamic programming embedded within each likelihood computation.

\subsection{Limitations of Existing MDP-Based Measurement Models} \label{subsec:limitations}

Although the MDP-MM links value-based decision models with latent-trait measurement, several aspects of its current formulation limit its applicability to more complex tasks. In particular,
\begin{enumerate}
    \item \emph{Scalability.}
    Computing $Q(s,a \mid \beta)$ in the MDP-MM requires solving the full MDP via dynamic programming over the task's state space. As \citet{lamar2018markov} notes, even in relatively simple applications the resulting state space can become large enough that exact dynamic programming is already demanding, which confines practical applications to small, fully enumerated environments.

    \item \emph{Coupling of $\beta_j$ and  $Q(s,a)$.}
    Because the value function $Q(s,a \mid \beta)$ is defined separately for each value of the ability parameter, the model tightly couples person parameters with the task representation. In numerical implementations, the value function must be recomputed for every quadrature node or trial value of $\beta$, which substantially increases the computational burden and makes it difficult to employ more flexible or high-dimensional parameterizations of $Q(s,a)$.

\item \emph{Fixed task model assumptions.}
In the MDP-MM, the transition model and reward parameters are treated as known and shared across examinees. This keeps the model simple by attributing individual differences primarily to $\beta_j$, but it also implies that differences in task understanding, goals, or strategy are not explicitly modeled and may be confounded with $\beta_j$ or reflected in poor fit. The assumption is plausible for simple, fully observable tasks (e.g., peg solitaire) but becomes more questionable as task structure becomes less transparent or state information becomes incomplete.

\end{enumerate}

Taken together, these limitations motivate measurement models that preserve the interpretability of MDP-based formulations while scaling more easily and allowing more flexible representations of task structure.

\section{Reinforcement Learning Measurement Model}
\label{sec:method}


As reviewed in Section~\ref{sec:background}, the MDP-MM specifies within-task action probabilities via the Boltzmann decision rule in Eq.~\eqref{eq:mdpmm}. For each value of the person parameter $\beta_j$, the corresponding action-value function $Q(s,a \mid \beta_j)$ is obtained by dynamic programming, and the distribution of $\beta_j$ is estimated by marginal maximum likelihood (Eq.~\eqref{eq:mdpmm_marglik}). This construction links value-based cognitive models and latent-trait measurement, but, as discussed in Section~\ref{subsec:limitations}, the reliance on tabular dynamic programming and $\beta$-specific value functions limits scalability and flexibility.

To address the limitations, we propose the Reinforcement Learning Measurement Model (RLMM), which can be viewed as a shared-value generalization of the MDP-MM. Both models use a Boltzmann choice rule in which the person parameter governs sensitivity to task-defined action scores, but they differ in how those scores are constructed. In the RLMM studied here, the tabular, person-specific value functions are replaced by a shared parametric value function $Q_\theta(s,a)$ estimated from observed trajectories, and the value function is constrained by a soft Bellman penalty that encodes the task dynamics.

\subsection{Model Specification}

We adopt the MDP notation from Section~\ref{sec:background}. Each task is represented as
\[
  (S,A,T,R,\gamma),
\]
where $S$ is the state space, $A$ the finite action set, $T(s' \mid s,a)$ the transition kernel, $R(s,a,s')$ the immediate reward, and $\gamma \in [0,1]$ the discount factor.

For participant $j=1,\dots,J$, we observe a sequence of state--action--next-state triples
\[
  \mathcal D_j = \{(s_{jt},a_{jt},s'_{jt}) : t=1,\dots,T_j\},
\]
and write $\mathcal D = \{\mathcal D_j\}_{j=1}^J$ for the full dataset. Our aim is to infer an interpretable ability parameter $\beta_j$ for each participant together with a shared value function $Q_\theta(s,a)$ that captures how actions are evaluated across states.

In place of the tabular, person-specific value functions $Q(s,a \mid \beta_j)$ in the MDP-MM, we introduce a single shared value function $Q_\theta(s,a)$, parameterized by $\theta$ (for example, via a neural network over state features). Because this shared-value parameterization separates the task-level score function from person-specific choice sensitivity, explicit identification constraints are needed. In particular, the softmax model in Eq.~\eqref{eq:softmax} is invariant to state-wise translation of the scores, and the pair $(Q_\theta,\beta_j)$ is behaviorally unchanged under global co-scaling. We therefore work with a centered and globally normalized score function constructed from $Q_\theta$.

For each state $s$, we first remove the state-wise mean,
\begin{equation}
  \tilde A_\theta(s,a)
  = Q_\theta(s,a) - \frac{1}{|A|} \sum_{a' \in A} Q_\theta(s,a'),
  \label{eq:adv_center}
\end{equation}
so that $\sum_{a \in A} \tilde A_\theta(s,a) = 0$. This centering step removes the state-wise translation freedom of the softmax policy. We then define a single global scale factor
\begin{equation}
  c_\theta
  =
  \left(
    \frac{1}{\sum_{j=1}^J T_j}
    \sum_{j=1}^J \sum_{t=1}^{T_j}
    \tilde A_\theta(s_{jt},a_{jt})^2
  \right)^{1/2},
  \label{eq:adv_scale}
\end{equation}
using the empirical distribution of observed state--action pairs, and normalize by this shared factor:
\begin{equation}
  A_\theta(s,a)
  = \frac{\tilde A_\theta(s,a)}{c_\theta},
  \label{eq:adv_norm}
\end{equation}
This construction fixes the overall behavioral scale of the score function while preserving relative differences in action gaps across states. Under a global rescaling $Q_\theta \mapsto c Q_\theta$ with $c>0$, both $\tilde A_\theta$ and $c_\theta$ scale by $c$, so $A_\theta$ is unchanged. Thus the policy in Eq.~\eqref{eq:policy} no longer admits the behavioral co-scaling freedom between $Q_\theta$ and $\beta_j$. In implementation, a small constant can be added inside the square root in Eq.~\eqref{eq:adv_scale} for numerical stability.

Given $A_\theta$, we define the decision model for participant $j$ by
\begin{equation}
  \pi_\theta(a \mid s,\beta_j)
  = \frac{\exp\{\beta_j A_\theta(s,a)\}}
  {\sum_{a' \in A} \exp\{\beta_j A_\theta(s,a')\}},
  \label{eq:policy}
\end{equation}
where $\beta_j>0$ is interpreted as an ability parameter: larger values of $\beta_j$ correspond to sharper preferences for higher-advantage actions.

The normalized policy defines local decision difficulty through the gaps among available actions at a state. States in which one action has a clear normalized advantage are easier under the model, and states in which the available actions have similar normalized advantages are harder. Suppose a state $s$ has a unique highest-advantage action
\[
  a_\theta^\star(s) = \arg\max_{a \in A} A_\theta(s,a).
\]
Then the probability that participant $j$ selects that optimal action can be written as
\begin{equation}
  p_{\mathrm{opt}}(s,\beta_j)
  = \pi_\theta(a_\theta^\star(s) \mid s,\beta_j)
  = \left[
      1 + \sum_{a \neq a_\theta^\star(s)}
      \exp\!\left\{
        -\beta_j \Delta_\theta(s,a)
      \right\}
    \right]^{-1},
  \label{eq:opt_prob}
\end{equation}
where
\[
  \Delta_\theta(s,a)
  = A_\theta\!\left(s,a_\theta^\star(s)\right) - A_\theta(s,a)
\]
is the advantage gap between the optimal action and a competing action. These gaps summarize the local separation among available actions. In a two-action state, this separation is represented by a single normalized gap, and Eq.~\eqref{eq:opt_prob} reduces to
\begin{equation}
  p_{\mathrm{opt}}(s,\beta_j)
  = \frac{1}{1 + \exp\!\left\{-\beta_j \Delta_\theta(s)\right\}},
  \label{eq:opt_prob_binary}
\end{equation}
where $\Delta_\theta(s)$ is the gap between the better and worse action at state $s$.

\begin{figure}[tbp]
  \centering
  \includegraphics[width=0.60\linewidth]{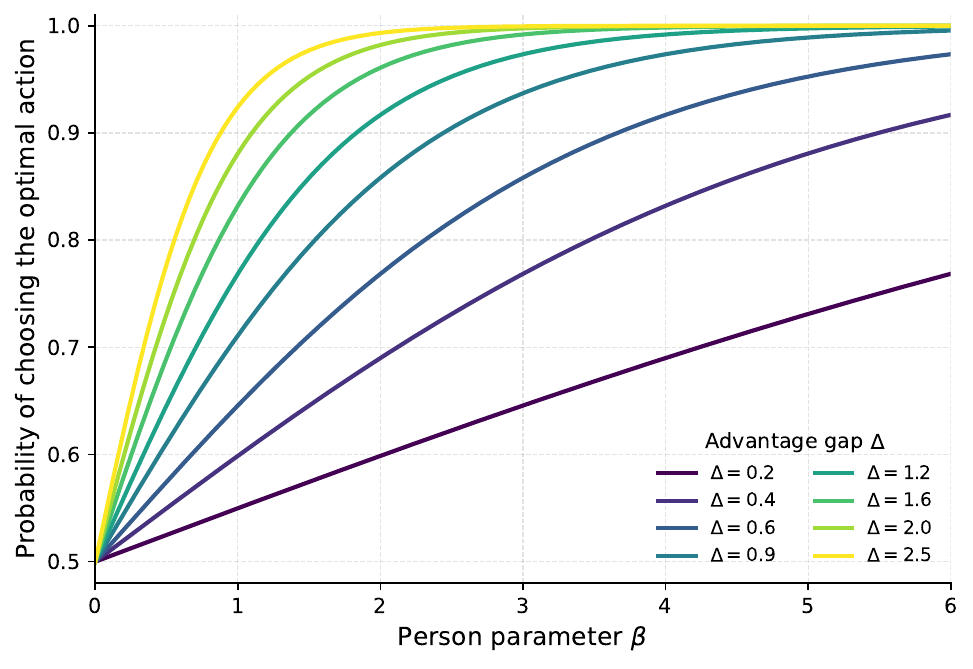}
  \caption{Binary decision characteristic curves for the optimal action as a function of the person parameter $\beta_j$, with separate curves for fixed normalized advantage gaps $\Delta_\theta(s)$}
  \label{fig:decision_characteristic_curves}
\end{figure}

Figure~\ref{fig:decision_characteristic_curves} plots Eq.~\eqref{eq:opt_prob_binary} for several values of $\Delta_\theta(s)$. For a fixed state, each curve corresponds to one normalized gap between the better and worse action. At the boundary value $\beta_j=0$, all curves equal $0.5$; as $\beta_j$ increases, the gap determines how quickly the optimal-action probability rises above chance. Large gaps produce steep curves and high optimal-action probabilities even at moderate values of $\beta_j$. Small gaps produce shallow curves, so the two actions remain weakly separated over a wider range of ability.

For regularization, and following \citet{lamar2018markov}, we place a log-normal prior on the ability parameters: 
\begin{equation}
\log \beta_j \sim \mathcal{N}(\mu,\sigma^{2}),
\qquad j=1,\ldots,J,
\label{eq:beta_prior}
\end{equation}
where $(\mu,\sigma^{2})$ characterize the population distribution of ability and the individual
$\beta_j$ are regularized toward this distribution.

\subsection{Penalized Likelihood and Estimation}

To connect the learned value function $Q_\theta$ to the known task dynamics $(T,R,\gamma)$, we impose a soft Bellman consistency penalty. We define a soft state-value function
\begin{equation}
  V_\theta(s)
  = \tau \log \sum_{a \in A} \exp\!\left\{\frac{1}{\tau} Q_\theta(s,a)\right\},
  \label{eq:soft_value}
\end{equation}
with a fixed temperature $\tau>0$ (e.g., $\tau=1$), and a corresponding Bellman residual
\begin{equation}
  \delta_\theta(s,a,s')
  = Q_\theta(s,a) - \bigl\{ R(s,a,s') + \gamma V_\theta(s') \bigr\}.
  \label{eq:bellman_resid}
\end{equation}
The Bellman term is defined as
\begin{equation}
  \mathcal L_{\text{bell}}(\theta)
  = \mathbb E_{(s,a,s') \sim \mathcal D}
    \bigl[ \delta_\theta(s,a,s')^2 \bigr],
  \label{eq:bellman_loss}
\end{equation}
that is, the mean squared Bellman residual under the empirical distribution of observed transitions. In the objective function (Eq.~\eqref{eq:pen_objective}), $\mathcal L_{\text{bell}}(\theta)$ enters with a weight $\lambda_{\text{bell}}>0$, so that estimation trades off goodness-of-fit to the observed actions against approximate satisfaction of the soft Bellman equation. When $\lambda_{\text{bell}}=0$ the value function is determined solely by the behavioral data, whereas larger values of $\lambda_{\text{bell}}$ place more emphasis on respecting the MDP dynamics in the sense of the soft Bellman residual.

Given the policy in Eq.~\eqref{eq:policy}, the negative log-likelihood of the observed action sequences is
\begin{equation}
  \mathcal L_{\text{beh}}(\theta,\{\beta_j\})
  = -\sum_{j=1}^J \sum_{(s,a) \in \mathcal D_j}
    \log \pi_\theta(a \mid s,\beta_j).
  \label{eq:behavior_nll}
\end{equation}
Because the prior in Eq.~\eqref{eq:beta_prior} is specified on the log-ability scale, we work with
\[
  z_j = \log \beta_j,
\]
so that $\beta_j = e^{z_j}$. Combining Eqs.~\eqref{eq:beta_prior} and \eqref{eq:bellman_loss}, we define the penalized objective
\begin{equation}
  \mathcal J(\theta,\{z_j\})
  = \mathcal L_{\text{beh}}(\theta,\{e^{z_j}\})
  + \lambda_{\text{bell}}\, \mathcal L_{\text{bell}}(\theta)
  + \sum_{j=1}^J \frac{(z_j-\mu)^2}{2\sigma^2},
  \label{eq:pen_objective}
\end{equation}
up to an additive constant that does not depend on $(\theta,\{z_j\})$, where $\lambda_{\text{bell}}>0$ controls the strength of the dynamic-consistency penalty and the final term is the negative log-prior for the log-ability parameters.

\begin{figure}[tbp]
  \centering
  \includegraphics[width=0.9\linewidth]{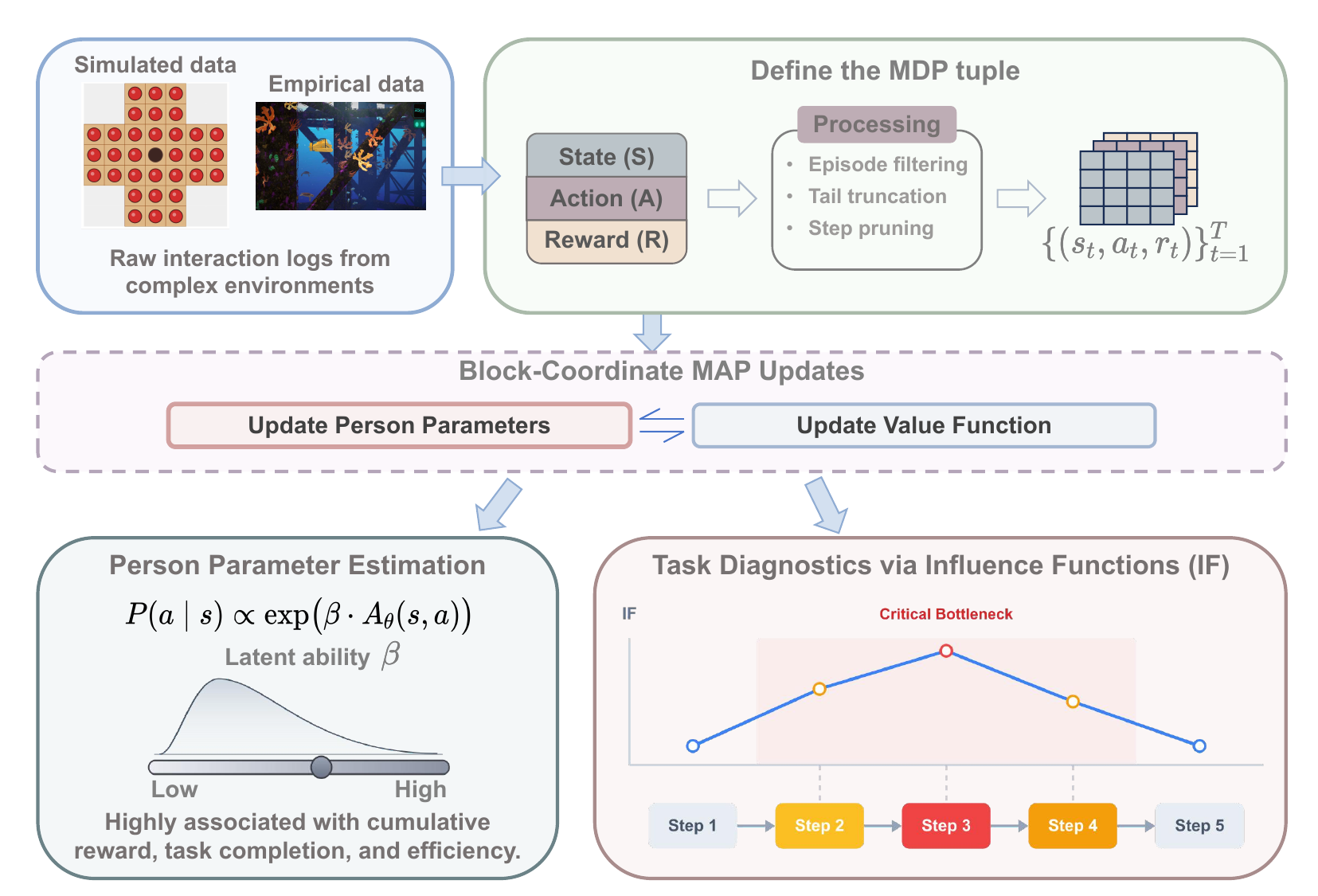}
  \caption{Overview of the RLMM workflow from interaction data to ability estimation and task diagnostics}
  \label{fig:rlmm_architecture}
\end{figure}

Direct joint minimization of $\mathcal J$ over $(\theta,\{z_j\})$ is challenging: the value-function parameters $\theta$ are of high dimension, and each log-ability parameter $z_j$ appears in many terms of the behavioral likelihood. To make estimation more tractable, we use a block-coordinate MAP procedure that alternates between updating the person parameters and updating the shared value function. In each outer iteration, we first fix $\theta$ and update the log-ability parameters $\{z_j\}$, and then fix $\{z_j\}$ and take a small number of stochastic gradient steps on $\theta$. This block-coordinate scheme is straightforward to implement and can accommodate large datasets; Figure~\ref{fig:rlmm_architecture} provides a visual overview of the estimation pipeline, and Algorithm~\ref{alg:block_coordinate_map} summarizes the procedure.

In the person-parameter update, we reparameterize the ability parameter via $z_j = \log \beta_j$. Conditional on the current value function, the log-posterior for $z_j$ is
\begin{equation}
  \ell_j(z)
  = \sum_{(s,a) \in \mathcal D_j}
      \Bigl[
        e^z A_\theta(s,a)
        - \log \sum_{a' \in A} \exp\{ e^z A_\theta(s,a') \}
      \Bigr]
    - \frac{(z-\mu)^2}{2\sigma^2},
  \label{eq:person_logpost}
\end{equation}
which is a smooth one-dimensional objective in $z$. We update $z_j$ by a small number of Newton--Raphson iterations
\[
  z^{(m+1)} = z^{(m)} - \frac{g_j(z^{(m)})}{H_j(z^{(m)})},
\]
where $g_j(z)$ and $H_j(z)$ denote the gradient and Hessian of $\ell_j(z)$, and then set $\hat\beta_j = \exp\{\hat z_j\}$. Because participants are conditionally independent given $\theta$, these one-dimensional updates are carried out separately for each $j$ and can be parallelized across persons.

In the value-function update, we hold $\{\hat z_j\}$ fixed, equivalently $\{\hat\beta_j\}$, and apply stochastic gradient descent to $\theta$ on the objective in Eq.~\eqref{eq:pen_objective},
\begin{equation}
  \theta \leftarrow \theta - \eta \, \nabla_\theta \mathcal J(\theta,\hat z),
  \label{eq:sgd_value_update}
\end{equation}
using mini-batches of transitions drawn from $\mathcal D$. In practice, we perform a fixed, small number of gradient steps per outer iteration rather than fully optimizing $\theta$ at each step. This makes the overall procedure computationally manageable in the applications we consider.

\begin{algorithm}[t]
  \caption{Block-Coordinate MAP Estimation for RLMM}
  \label{alg:block_coordinate_map}
  \begin{algorithmic}[1]
    \State Initialize value-function parameters $\theta^{(0)}$
           and person parameters $\{\beta_j^{(0)}\}_{j=1}^J$.
    \For{$k = 0,1,2,\dots$ until convergence}
      \State \textbf{Person-parameter update}
      \For{each participant $j = 1,\dots,J$}
        \State Set $z_j^{(0)} = \log \beta_j^{(k)}$.
        \For{$m = 0,1,\dots,M_{\text{NR}}-1$}
          \State Compute gradient $g_j(z_j^{(m)})$
                 and Hessian $H_j(z_j^{(m)})$ of $\ell_j(z)$
                 in Eq.~\eqref{eq:person_logpost}.
          \State Update
            $z_j^{(m+1)} = z_j^{(m)} - g_j(z_j^{(m)}) / H_j(z_j^{(m)})$.
        \EndFor
        \State Set $z_j^{(k+1)} = z_j^{(M_{\text{NR}})}$ and
               $\beta_j^{(k+1)} = \exp\{ z_j^{(k+1)} \}$.
      \EndFor
      \State \textbf{Value-function update}
      \For{$m = 1,\dots,M_{\text{SGD}}$}
        \State Sample a mini-batch of transitions from $\mathcal D$.
        \State Compute a stochastic gradient
               $\nabla_\theta \mathcal J(\theta^{(k)},\{z_j^{(k+1)}\})$
               on the mini-batch.
        \State Update
          $\theta^{(k)} \leftarrow \theta^{(k)}
            - \eta \, \nabla_\theta \mathcal J(\theta^{(k)},\{z_j^{(k+1)}\})$.
      \EndFor
      \State Set $\theta^{(k+1)}$ to the updated value of $\theta^{(k)}$.
    \EndFor
  \end{algorithmic}
\end{algorithm}

\subsection{Influence-Function Diagnostics for Critical Steps}
\label{subsec:influence}

Beyond estimating a global ability parameter $\beta_j$, we also aim to identify which specific decisions provide the strongest evidence for or against a participant's estimated ability. We use an influence-function approximation that quantifies how much the person estimate would change if a single step were given slightly more weight in the person-specific objective.

Fix a participant $j$ and work with $z_j=\log\beta_j$ as in the person-parameter update. Let $\ell_j(z)$ denote the person-specific log-posterior in Eq.~\eqref{eq:person_logpost}, and let $\ell_{jt}(z)$ be the step-wise log-likelihood contribution at time $t$. Consider an $\varepsilon$-perturbed objective that upweights step $t$,
\begin{equation}
  \ell_j^{(\varepsilon,t)}(z)
  = \ell_j(z) + \varepsilon\, \ell_{jt}(z),
  \label{eq:perturbed_obj}
\end{equation}
and let $\hat z_j(\varepsilon)$ be its maximizer. The influence of step $t$ on the estimate $\hat z_j$ is
\begin{equation}
  \mathcal I_{jt}
  \triangleq
  \left.\frac{d\,\hat z_j(\varepsilon)}{d\varepsilon}\right|_{\varepsilon=0}
  \approx
  -\Bigl[\nabla_z^2 \ell_j(\hat z_j)\Bigr]^{-1}\,
   \nabla_z \ell_{jt}(\hat z_j),
  \label{eq:if_def}
\end{equation}
where the Hessian term is a scalar because $z$ is one-dimensional. In our implementation, $\nabla_z^2 \ell_j(\hat z_j)$ is available from the Newton--Raphson updates used to obtain $\hat z_j$.

Using the Boltzmann policy in Eq.~\eqref{eq:policy} and the reparameterization $\beta=e^z$, the step score admits the closed form
\begin{equation}
  \nabla_z \ell_{jt}(z)
  = \beta\!\left(
      A_\theta(s_{jt},a_{jt})
      - \mathbb E_{a \sim \pi_\theta(\cdot \mid s_{jt},\beta)}
        \bigl[A_\theta(s_{jt},a)\bigr]
    \right).
  \label{eq:if_score}
\end{equation}
The step score can also be expressed in terms of the action-gap notation introduced above. Let the policy-weighted mean gap at state $s$ be
\begin{equation}
  \bar\Delta_\pi(s,\beta)
  =
  \sum_{a \in A}
  \pi_\theta(a \mid s,\beta)\Delta_\theta(s,a).
  \label{eq:mean_gap}
\end{equation}
Since $A_\theta(s,a)=A_\theta(s,a_\theta^\star(s))-\Delta_\theta(s,a)$, Eq.~\eqref{eq:if_score} can be rewritten as
\begin{equation}
  \nabla_z \ell_{jt}(z)
  =
  \beta\!\left\{
    \bar\Delta_\pi(s_{jt},\beta)
    -
    \Delta_\theta(s_{jt},a_{jt})
  \right\}.
  \label{eq:if_gap_score}
\end{equation}
This form shows how a decision step affects the person estimate. It compares the policy-weighted gap at the current state with the gap of the action actually chosen. The state-level gap structure therefore contributes to the influence score, but its effect depends on the realized action. Choosing the highest-advantage action yields positive evidence for a larger $\beta_j$, whereas choosing a lower-advantage action yields negative evidence. In the two-action case, Eq.~\eqref{eq:if_gap_score} reduces to
\begin{equation}
  \nabla_z \ell_{jt}(z)
  =
  \begin{cases}
    \beta\{1-p_{\mathrm{opt}}(s_{jt},\beta)\}\Delta_\theta(s_{jt}),
    & a_{jt}=a_\theta^\star(s_{jt}),\\
    -\beta p_{\mathrm{opt}}(s_{jt},\beta)\Delta_\theta(s_{jt}),
    & a_{jt}\neq a_\theta^\star(s_{jt}).
  \end{cases}
  \label{eq:if_score_binary}
\end{equation}
Equations~\eqref{eq:if_gap_score} and \eqref{eq:if_score_binary} show where local decision difficulty enters the influence calculation. The action gaps determine the step-score contribution, and the chosen action determines whether that contribution supports a larger or smaller value of $\beta_j$. The curvature term in Eq.~\eqref{eq:if_def} then maps this score contribution to the local perturbation $\mathcal I_{jt}$ on the log-ability scale used for estimation.

We interpret the influence score $\mathcal I_{jt}$ as the local contribution of step $t$ for person $j$ to the ability estimate. Its sign indicates direction: $\mathcal I_{jt}>0$ provides evidence for higher ability, whereas $\mathcal I_{jt}<0$ provides evidence for lower ability. To identify critical steps regardless of direction, we rank steps by the magnitude $|\mathcal I_{jt}|$. We summarize influential steps using reporting rules (e.g., top-$K$ steps or percentile cutoffs) described in the experimental section.

\subsection{Relationship to the MDP Measurement Model and Scalability Considerations}
\label{subsec:mdpmm_scaling}

The RLMM can be viewed as a shared-value generalization of the original MDP-MM. In both models, the person parameter governs sensitivity to task-defined action scores. The difference is that the MDP-MM uses person-specific tabular value functions, whereas the present RLMM uses normalized advantages $A_\theta(s,a)$ derived from a shared parametric value function and regularized by the Bellman penalty in Eq.~\eqref{eq:bellman_loss}. This reformulation replaces ability-specific tabular value tables with a single learned representation and leads to the computational savings described below.

The key computational cost in the MDP-MM is that value functions are computed by tabular dynamic programming for each quadrature node over $\beta$ in the marginal likelihood. Let $|S|$ and $|A|$ denote the numbers of states and actions, $N_Q$ the number of quadrature nodes, and $N_{\mathrm{PI}}$ the number of outer policy-iteration iterations. In a dense-transition worst case, policy evaluation requires solving the Bellman system
$(I-\gamma P^\pi)V^\pi=r^\pi$,
which scales as $O(|S|^3)$ per iteration under direct linear solves, yielding an overall cost on the order of
\begin{equation}
  O\!\left(N_Q\, N_{\mathrm{PI}}\, |S|^3\right).
  \label{eq:mdpmm_complexity_dense}
\end{equation}

Even under more favorable sparsity (e.g., deterministic transitions), tabular dynamic programming still entails repeated Bellman updates over the entire enumerated state space (and typically all state--action pairs).

In contrast, we learn a shared parametric value function $Q_\theta(s,a)$ with $P$ trainable parameters from logged trajectories. Let $T=\sum_{j=1}^J T_j$ be the total number of observed transitions, $B$ the mini-batch size, and suppose we perform one gradient update every $k$ transitions (up to constant factors, this matches standard mini-batch stochastic optimization). Each transition contributes $O(P)$ work for forward or backward passes through $Q_\theta$, so the total cost of fitting $Q_\theta$ scales approximately as
\begin{equation}
  O(TP),
  \label{eq:rl_complexity}
\end{equation}
for fixed architecture (fixed $P$). Person-parameter updates are one-dimensional. With $M_{\mathrm{NR}}$ Newton--Raphson steps and a softmax normalization over the available actions, the person-parameter update cost is
\begin{equation}
  O\!\left(M_{\mathrm{NR}} \sum_{j=1}^J \sum_{t=1}^{T_j} |A(s_{jt})|\right),
  \label{eq:person_update_complexity}
\end{equation}
which is linear in the amount of logged data and does not require dynamic-programming updates over all enumerated states. As summarized in Table~\ref{tab:complexity}, the MDP-MM incurs an explicit dependence on $|S|$ in its dynamic-programming stage, and this cost is further multiplied by $N_Q$ because the value function must be recomputed across quadrature nodes for $\beta$.

\begin{table}[t]
\centering
\caption{Computational scaling comparison (dominant terms)}
\label{tab:complexity}
\begin{tabular}{lcc}
\hline
Component & MDP-MM & RLMM \\
\hline
Value-function fit
& $O(N_Q N_{\mathrm{PI}}\, \cS{|S|^3})$
& $O(TP)$ \\
Value-function memory
& $O(\cS{|S|}\,|A|)$
& $O(P)$ \\
\hline
\end{tabular}
\end{table}

These differences matter even for moderately large discrete tasks. For a grid world with $100\times 100$ locations, the state space already has $|S|\approx 10^4$ states, and realistic variants with obstacles, objects, or richer observations can easily reach $10^5$--$10^6$ distinct states. Applying the MDP-MM would require repeated tabular dynamic-programming solves over all state--action pairs for each quadrature node, which quickly becomes infeasible. The RLMM can instead represent $Q_\theta(s,a)$ using state features (e.g., convolutional features over grid layouts) and fit it from logged trajectories with mini-batch optimization, while still producing person-level ability parameters.

Similar scaling issues arise in adaptive educational systems. In that setting, the state summarizes a student's response history and timing, and the actions correspond to selecting items or learning activities from a large pool. If each distinct history is treated as a separate state, the resulting discrete state space is effectively unbounded, which makes tabular dynamic programming impractical. A parameterized $Q_\theta(s,a)$ can generalize across histories and items, so the framework retains the same measurement logic, namely a Boltzmann policy with a log-normal ability parameter, while remaining trainable in large logged-data settings where full enumeration of $S$ is not possible.

\section{Simulation Studies} \label{sec:experiments}

We evaluate the Reinforcement Learning Measurement Model (RLMM) on peg-solitaire tasks of increasing complexity. The first set of experiments revisits the four boards analyzed by \citet{lamar2018markov} and compares the RLMM with the original MDP-MM under closely matched conditions. The second set introduces two larger boards for which the tabular dynamic-programming approach underlying the MDP-MM is no longer computationally practical, and uses these tasks to assess the scalability and stability of the RLMM.

\subsection{Original peg-solitaire tasks}
\label{subsec:peg_tasks}

\begin{figure}[tbp]
  \centering
  \includegraphics[width=0.6\linewidth]{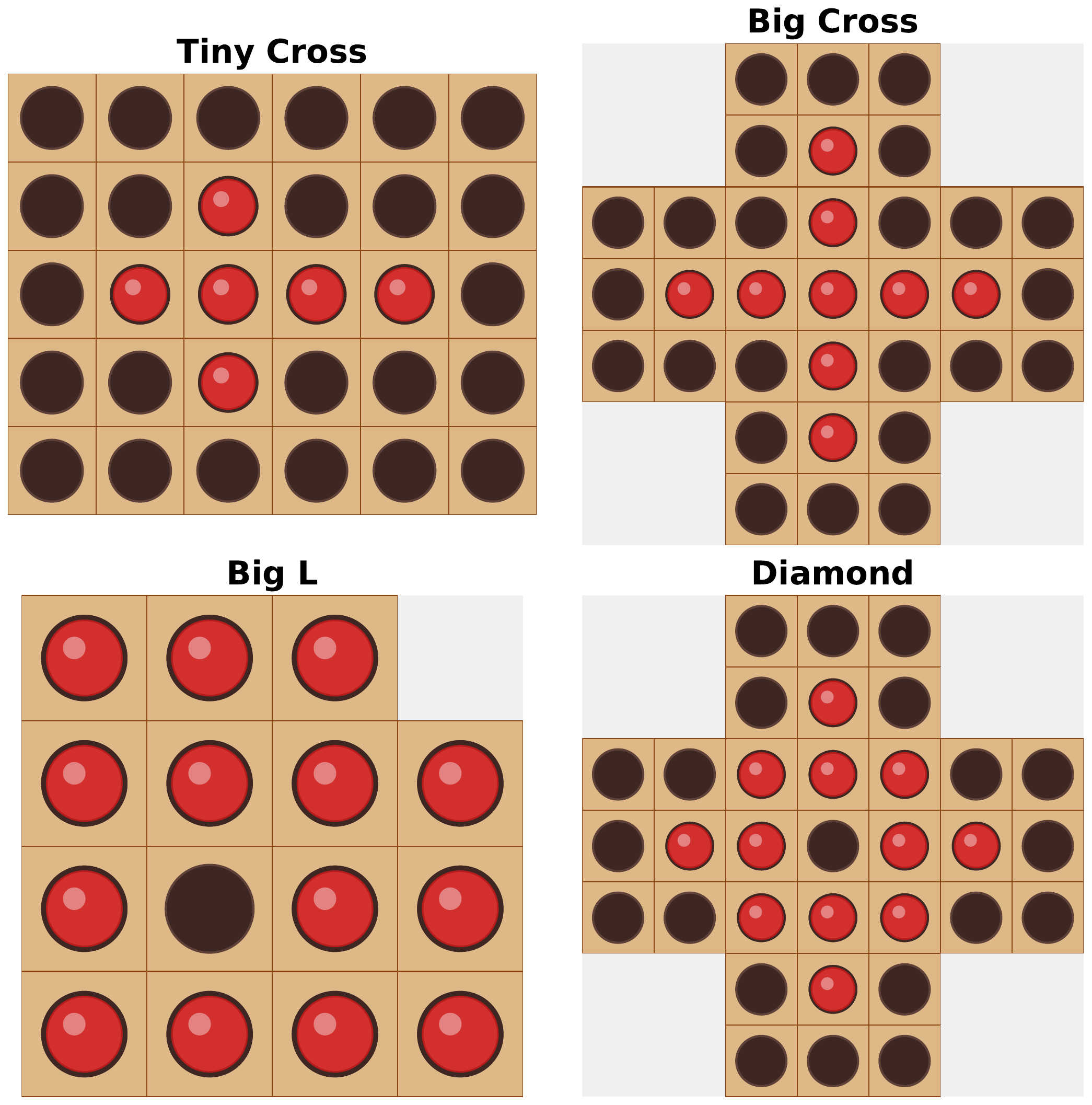}
  \caption{Layouts of the four peg-solitaire boards: tiny cross, big cross, Big-L, and diamond}
  \label{fig:peg_four_boards}
\end{figure}

The first study follows the peg-solitaire application in \citet{lamar2018markov}. We consider the same four boards, which differ in layout, solution-path length, and state-space size: a tiny cross board, a larger cross, an L-shaped board, and a diamond-shaped board. Figure~\ref{fig:peg_four_boards} illustrates the layouts of these tasks, and Table~\ref{tab:board_lamar} summarizes their basic characteristics. For each board we enumerate the reachable state space under legal moves and all legal actions, and we adopt the same transition and reward specification as in \citet{lamar2018markov}.

\begin{table}[t]
  \centering
    \caption{Task characteristics for the four peg-solitaire boards from \citet{lamar2018markov}}
  \label{tab:board_lamar}
  \begin{tabular}{lcccc}
    \toprule
    Board name  & Solution path length & $|S|$ (states) & $|A|$ (actions) \\
    \midrule
    Tiny cross  & 5   & 22    & 12 \\
    Big cross   & 8   & 153   & 22 \\
    Big-L       & 13  & 807   & 30 \\
    Diamond     & 11  & 5{,}923 & 70 \\
    \bottomrule
  \end{tabular}
\end{table}

To study measurement properties when true ability parameters are known, we adopt the simulation design of \citet{lamar2018markov}. For each of $J=200$ simulated students, an ability parameter $\beta_j$ is drawn from the log-normal distribution in Eq.~\eqref{eq:beta_prior}, with $(\mu,\sigma^2)$ set to match \citet{lamar2018markov}. Each student completes $50$ games on each of the four boards, and actions are generated from the MDP-MM with the softmax decision rule in Eq.~\eqref{eq:mdpmm}, producing $4{,}000$ games in total and the transition data $\mathcal D = \{\mathcal D_j\}_{j=1}^J$.

We then fit the RLMM from Section~\ref{sec:method} to this simulated dataset. The model specifies a single shared value function $Q_\theta(s,a)$ across students, together with person-specific ability parameters $\beta_j$ that enter through the normalized advantage function $A_\theta$ in Eq.~\eqref{eq:adv_norm}. Estimation follows the block-coordinate MAP procedure in Algorithm~\ref{alg:block_coordinate_map}, with the Bellman penalty weight $\lambda_{\text{bell}}$ fixed in advance and the parameters $(\mu,\sigma^2)$, $\theta$, and $\{\beta_j\}$ estimated from the data. We evaluate ability-estimation accuracy and predictive performance of the RLMM by comparing its results with the MDP-MM recovery statistics reported in \citet{lamar2018markov}.

We assess the quality of ability estimation by comparing the estimated ability parameters with the true values used in data generation. For each student $j$ and for each model $m \in \{\text{MDP-MM}, \text{RLMM}\}$, we obtain an estimate $\hat{\beta}^{(m)}_j$. As a scalar summary of recovery accuracy, we compute, for each board and each model, the root mean squared error (RMSE) of $\log \beta_j$,
\[
  \mathrm{RMSE}^{(m)}
  = \sqrt{\frac{1}{J} \sum_{j=1}^J
      \bigl( \log \hat{\beta}^{(m)}_j - \log \beta_j \bigr)^2 }.
\]

To further examine recovery for the RLMM, we also inspect scatterplots of true versus estimated $\log \beta_j$, which we use to assess overall fit and to check for systematic deviations from the diagonal.

\begin{figure}[tbp]
  \centering
  \includegraphics[width=0.9\linewidth]{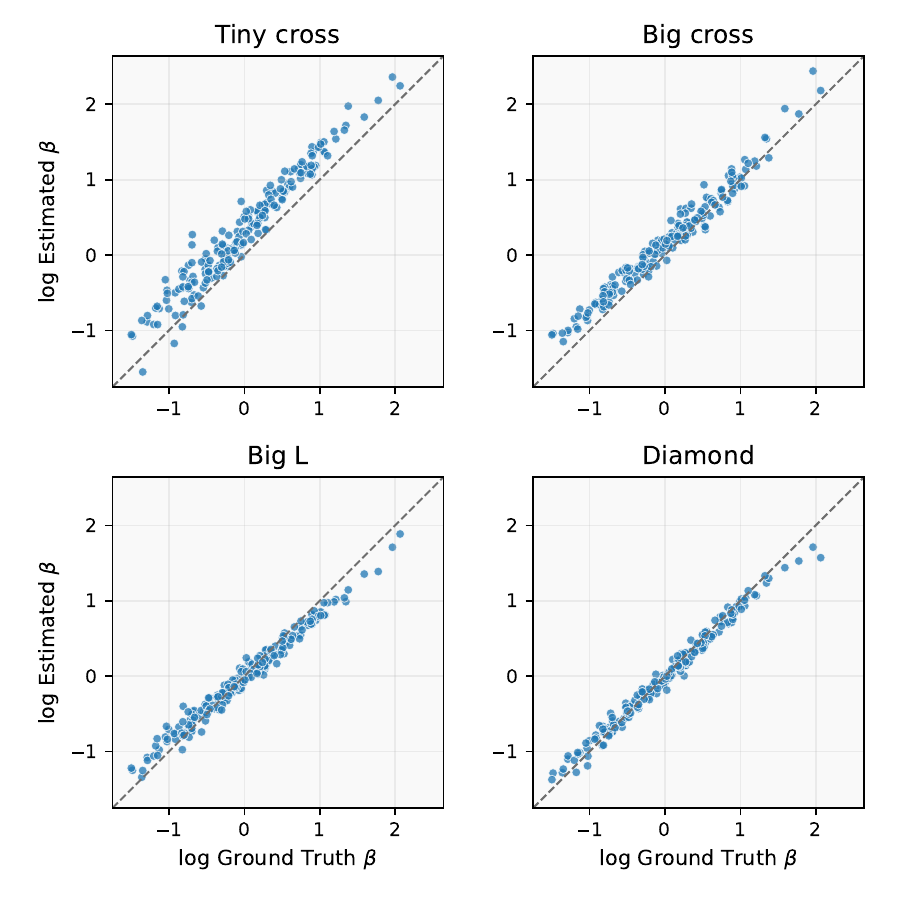}
 \caption{True versus estimated $\log \beta_j$ for the RLMM across four peg-solitaire boards (Tiny cross, Big cross, Big-L, and Diamond). Each point corresponds to one simulated participant. The dashed diagonal line indicates perfect recovery}
  \label{fig:beta_scatter_lamar}
\end{figure}

Table~\ref{tab:rmse_lamar} reports the RMSE of $\log \beta_j$ for the MDP-MM and the RLMM on each board. Across all four tasks, the RLMM attains lower RMSE than the MDP-MM, with relative reductions in error from roughly one-third on the smallest board to more than two-thirds on the diamond board. Figure~\ref{fig:beta_scatter_lamar} plots true against estimated $\log \beta_j$ for the RLMM across four peg-solitaire boards. In all cases, estimates lie close to the diagonal, indicating good recovery of the person parameters.

\begin{table}[t]
  \centering
  \caption{Comparison of RMSE for $\log \beta_j$ on the four peg-solitaire tasks from \citet{lamar2018markov}, reporting results for both the MDP-MM and the RLMM}

  \label{tab:rmse_lamar}
  \begin{tabular}{lcc}
    \toprule
    Board      & RMSE (MDP-MM) & RMSE (RLMM) \\
    \midrule
    Tiny cross & 0.542 & 0.368 \\
    Big cross  & 0.442 & 0.203 \\
    Big-L      & 0.388 & 0.142 \\
    Diamond    & 0.403 & 0.099 \\
    \bottomrule
  \end{tabular}
\end{table}

These results indicate that, on the original peg-solitaire tasks for which the MDP-MM was developed, the RLMM preserves the measurement structure of the MDP-MM while providing more accurate ability estimates and comparable or better predictive performance.

\subsection{Computational Scalability}
\label{subsec:scalability}

In addition to parameter recovery, we compare the computational efficiency of the RLMM estimation procedure with the original MDP-MM. We focus on the runtime of the second stage of estimation, which dominates the overall computational cost in both approaches. Table~\ref{tab:runtime_comparison} reports the wall-clock time for each method across the four peg-solitaire boards.

\begin{table}[t]
  \centering
  \caption{Runtime (in seconds) comparison between MDP-MM \citep{lamar2018markov} and the RLMM}
  \label{tab:runtime_comparison}
  \begin{tabular}{lccc}
    \toprule
    & MDP-MM  & RLMM  & Speedup \\
    \midrule
    Tiny cross  & 35.30  & 16.45 & 2.1$\times$ \\
    Big cross   & 75.52  & 17.12 & 4.4$\times$ \\
    Big-L       & 202.48 & 18.89 & 10.7$\times$ \\
    Diamond     & 370.53 & 19.35 & 19.1$\times$ \\
    \bottomrule
  \end{tabular}
\end{table}

For the smallest board (tiny cross), where the reachable state space is limited, the baseline MDP-MM remains computationally feasible, although the RLMM already achieves a noticeable speedup. As the board complexity increases, however, the computational burden of the MDP-MM grows rapidly. In particular, moving from the tiny cross to the diamond board increases the state space by more than two orders of magnitude (Table~\ref{tab:board_lamar}), and the corresponding runtime of the MDP-MM increases by an order of magnitude.

In contrast, the runtime of the RLMM remains stable across all four environments, staying within a narrow range of approximately 15--20 seconds. The runtime remains stable because the estimation procedure avoids repeatedly solving the full MDP during parameter updates. A shared value-function representation is learned across students, which limits the growth of computational cost as the state space expands.

Overall, these results indicate that the RLMM scales substantially better than the original MDP-MM as task complexity increases, while maintaining comparable or improved ability-estimation accuracy.

\subsection{Influence-Function Diagnostics}
\label{subsec:influence_difficulty}

Beyond global ability estimation, we use influence functions to diagnose which decision steps are most critical for successful performance, and which participants are most affected by errors at those steps. We focus on the Big-L board, which admits long solution paths and substantial branching early in the task.

Let $N(s)$ denote the number of solution paths from state $s$ that can still reach a successful terminal configuration. Averaged over trajectories starting from the initial state, $N(s)$ exhibits a characteristic collapse pattern. Around step 3, $N(s)$ typically shows its first sharp decrease, indicating that many locally plausible actions already eliminate large portions of the solution space. By steps 5--6, $N(s)$ often contracts from dozens or hundreds of remaining solutions to only a few viable paths. By step 8, the task frequently enters a narrow regime in which the state is effectively either solvable or irrecoverably failed.

This pattern implies that errors in steps 3--8 are rarely superficial. The most consequential mistakes are not merely suboptimal local gains, but actions that remove structural resources required for later connectivity, such as moves that block access to corners or disconnect regions of the board. Once such resources are lost, no subsequent actions can repair the resulting dead states.

\begin{figure}[tbp]
  \centering
  \includegraphics[width=0.85\linewidth]{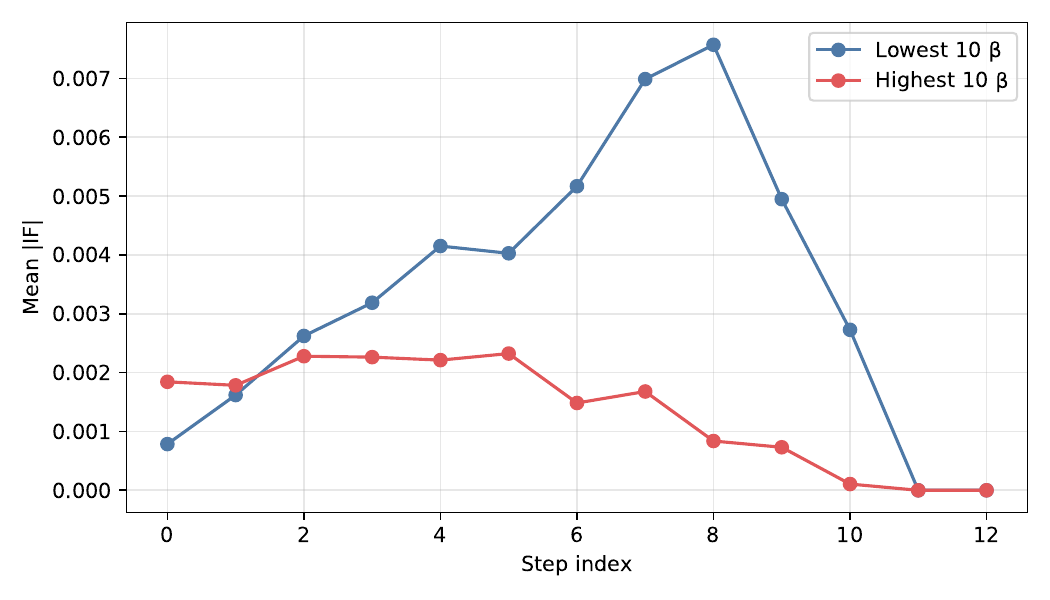}
  \caption{Mean absolute influence by step index for the lowest and highest $\beta$ participants on the Big-L board}
  \label{fig:top_bottom_if}
\end{figure}

Figure~\ref{fig:top_bottom_if} reports the mean absolute influence values across steps for the lowest- and highest-$\beta$ participants. For steps 3--8, the lowest-$\beta$ group exhibits substantially larger absolute influence values than both the highest-$\beta$ group and the remaining participants. In contrast, the highest-$\beta$ participants show comparatively small influence magnitudes in this interval, indicating that their decisions at these steps tend to be more robust with respect to the final outcome.

This contrast supports the interpretation that steps 3--8 constitute the most difficult region of the task. When decisions are made correctly, their marginal impact on the estimated ability is limited; when mistakes occur, however, they exert a disproportionately large negative influence on the likelihood and on the resulting $\beta$ estimates. Low-$\beta$ participants are more likely to incur such high-impact errors in this region, leading to both poorer outcomes and larger influence magnitudes.

\begin{figure}[tbp]
  \centering
  \includegraphics[width=0.9\linewidth]{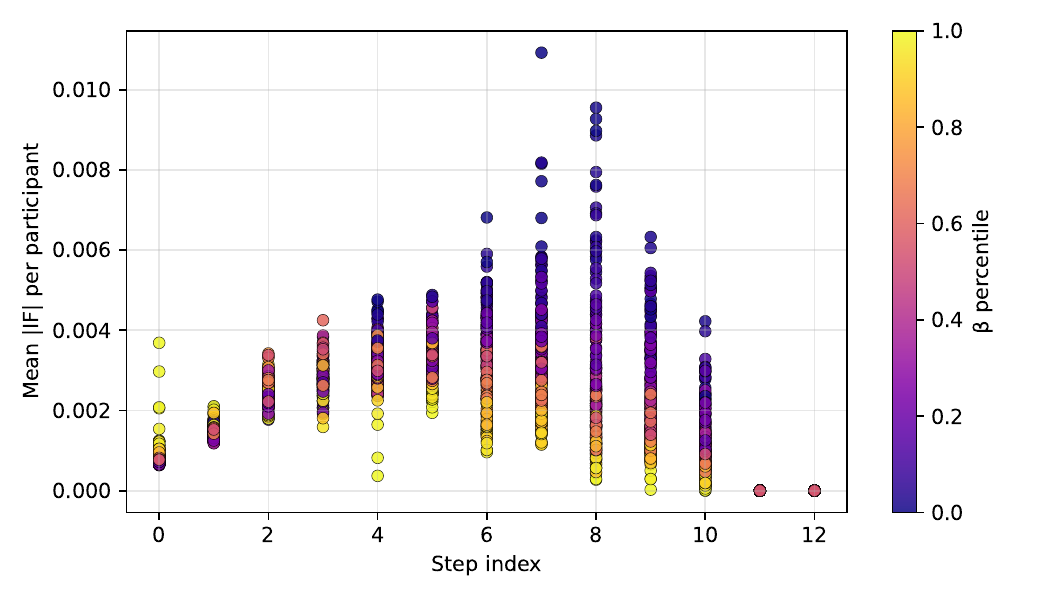}
  \caption{Scatter plot of mean absolute influence per participant by step index, colored by $\beta$ percentile}
  \label{fig:scatter_if}
\end{figure}

The same pattern is visible at the individual level in Figure~\ref{fig:scatter_if}. Participants with higher estimated ability tend to have uniformly small absolute influence values in steps 3--8, suggesting stable decision making under structural constraints. In contrast, lower-performing participants show both higher influence magnitudes and greater variability across these steps, reflecting sensitivity to local errors precisely where the task affords little tolerance for mistakes.

Taken together, these results show that influence functions provide a principled diagnostic tool for identifying both structurally difficult steps in a task and the participants for whom those steps are most consequential. In the Big-L board, steps 3--8 emerge as the primary bottleneck, and the concentration of large influence values among the lowest-$\beta$ participants explains their disproportionate performance degradation in this region.

\subsection{Larger peg-solitaire tasks}

The second study examines peg-solitaire tasks that exceed the practical range of the tabular MDP-MM. We construct two boards that are structurally similar to the Lamar boards but substantially larger in their state spaces and branching factors. The first is a $4\times 4$ grid with pegs on all cells except one initial empty position. The second is a $7\times 7$ cross-shaped board analogous to the standard English peg-solitaire layout. For each board we enumerate the reachable state space under legal moves and compute the shortest solution path length.

\begin{figure}[tbp]
  \centering
  \includegraphics[width=0.6\linewidth]{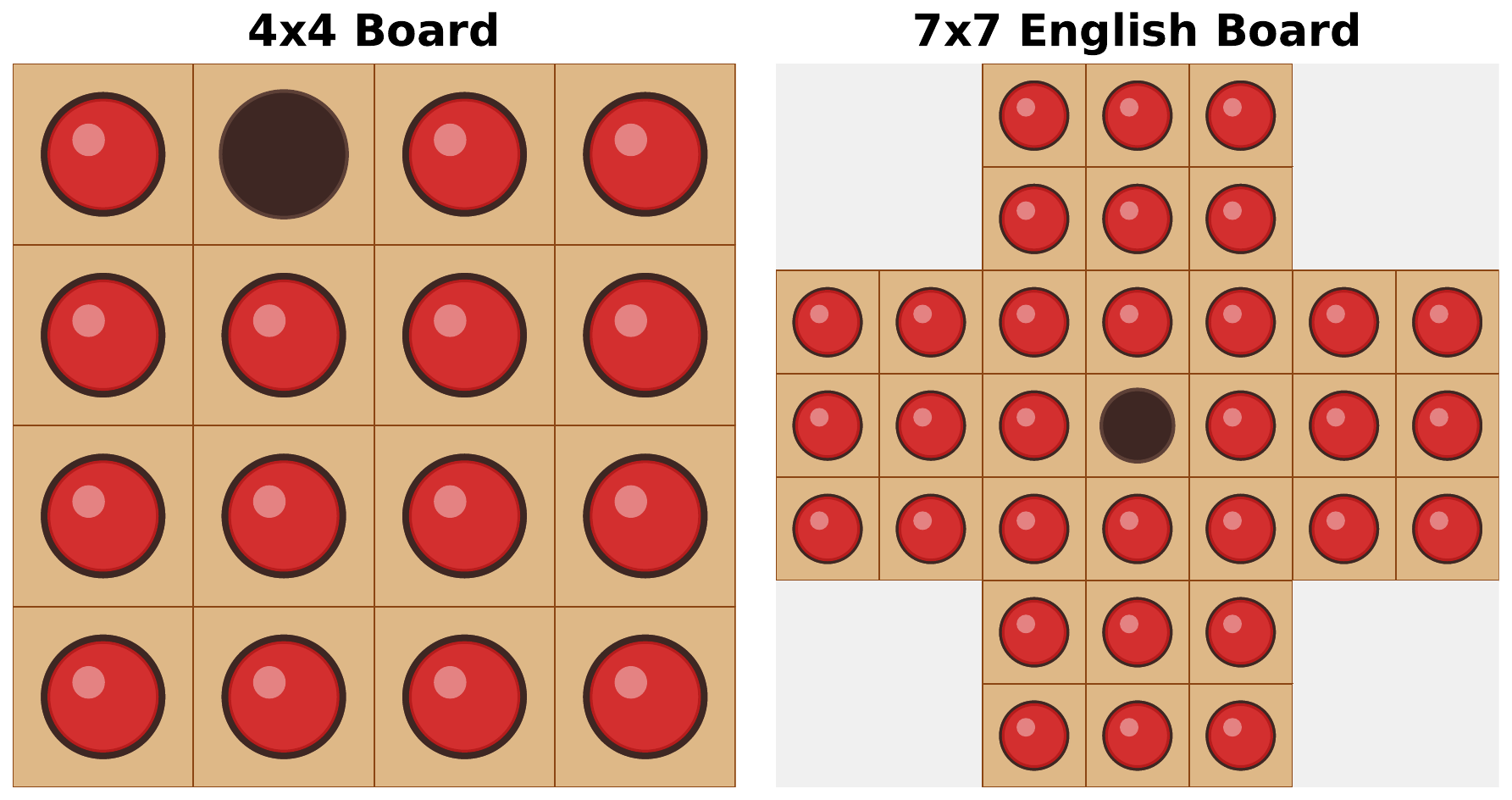}
  \caption{Layouts of the two larger peg-solitaire boards used in the scalability study (4$\times$4 grid and 7$\times$7 cross)}
  \label{fig:peg_large_boards}
\end{figure}

Table~\ref{tab:board_large} summarizes the main characteristics of the two larger tasks. The $4\times 4$ board has over $9{,}000$ reachable states, and the $7\times 7$ cross has more than $23$ million states; for comparison, the largest Lamar board (the diamond) has $5{,}923$ states. Figure~\ref{fig:peg_large_boards} shows the layouts of these boards. In practice, state spaces on the order of $5\times 10^5$ states are already near the practical limit for tabular dynamic programming combined with quadrature-based marginal likelihood in the MDP-MM. On the $7\times 7$ board, a single tabular value function with one entry per state--action pair would contain roughly $1.8$ billion values, requiring tens of gigabytes of memory in double precision, and recomputing such tables for each quadrature node or trial value of $\beta$ would be prohibitively expensive. For these reasons we do not attempt to fit the original MDP-MM on the larger boards.

\begin{table}[t]
  \centering
\caption{Task characteristics for the two larger peg-solitaire boards}
  \label{tab:board_large}
  \begin{tabular}{lcccc}
    \toprule
    Board name & Solution path length & $|S|$ (states) & $|A|$ (actions) \\
    \midrule
    $4\times 4$ grid   & 14 & 9{,}336       & 32 \\
    $7\times 7$ cross  & 23 & 23{,}475{,}688 & 76 \\
    \bottomrule
  \end{tabular}
\end{table}

Since the exact dynamic programming required for MDP-MM becomes infeasible at scale, we restrict our analysis of the larger boards to the RLMM. We computed the RMSE of $\log \beta_j$ using the same ground-truth comparison methodology detailed in Section~\ref{subsec:peg_tasks}.

\begin{figure}[tbp]
  \centering
  \includegraphics[width=0.9\linewidth]{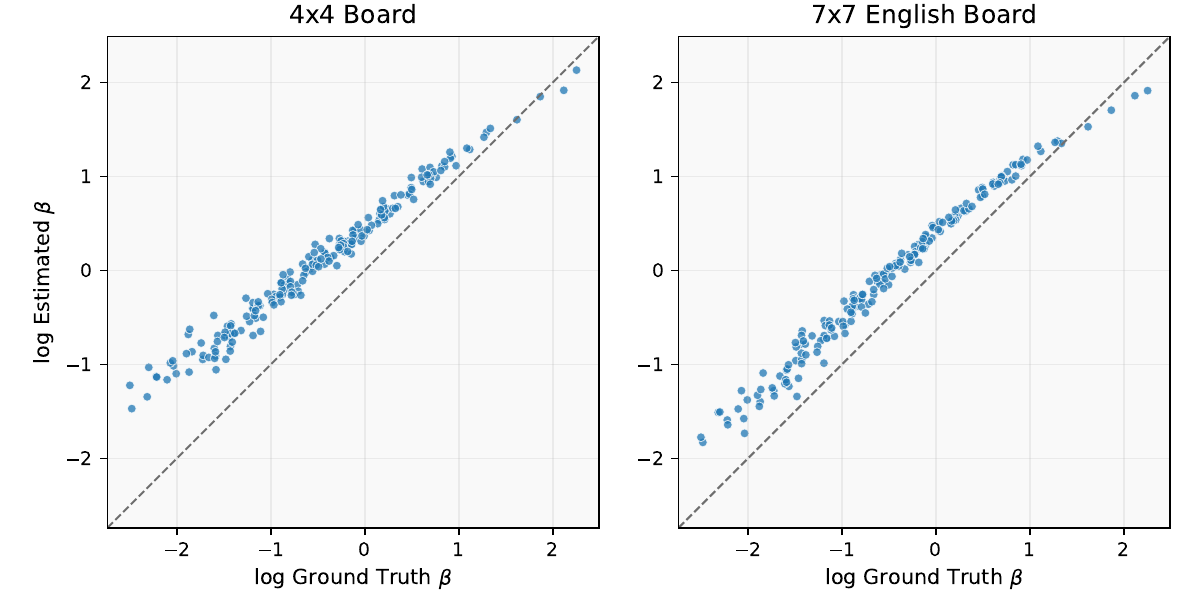}
  \caption{True versus estimated $\log \beta_j$ for the RLMM on the two larger boards ($4\times4$ grid and $7\times7$ English cross). Each point is one simulated participant. The dashed line indicates perfect recovery}
  \label{fig:beta_recovery_large}
\end{figure}

As shown in Table~\ref{tab:rmse_large}, the RLMM successfully recovers ability parameters despite the complexity of these larger environments. Figure~\ref{fig:beta_recovery_large} plots the estimated versus true $\log \beta_j$, revealing a strong correlation centered on the diagonal.

\begin{table}[t]
  \centering
  \caption{RMSE of $\log \beta_j$ for the RLMM on the two larger peg-solitaire boards}
  \label{tab:rmse_large}
  \begin{tabular}{lc}
    \toprule
    Board & RMSE (RLMM) \\
    \midrule
    $4\times 4$ grid   & 0.623 \\
    $7\times 7$ cross  & 0.452 \\
    \bottomrule
  \end{tabular}
\end{table}

\section{Empirical Study: AQUALAB}
\label{sec:empirical}
We apply the Reinforcement Learning Measurement Model (RLMM) to gameplay logs from Wake: Tales from the Aqualab (AQUALAB), an open-ended educational science game centered on the investigation of real-world marine ecosystems. The empirical analysis uses the February 2026 AQUALAB log export released through Open Game Data~\citep{FieldDayLabAqualabData}. A broader description of the Open Game Data infrastructure is provided by \citet{GagnonSwanson2023OpenGameData}. In contrast to the peg-solitaire tasks in Section~\ref{sec:experiments}, AQUALAB does not admit a small, fully enumerated state space with a fixed tabular solution. Accordingly, the purpose of this empirical study is not to evaluate parameter recovery against known ground truth, but to examine whether the estimated person parameter is associated with observable markers of successful task progression in a realistic and behaviorally noisy environment.

\subsection{Data and Preprocessing}
\label{subsec:aqualab_data}

\begin{figure}[tbp]
  \centering
  \includegraphics[width=0.95\linewidth]{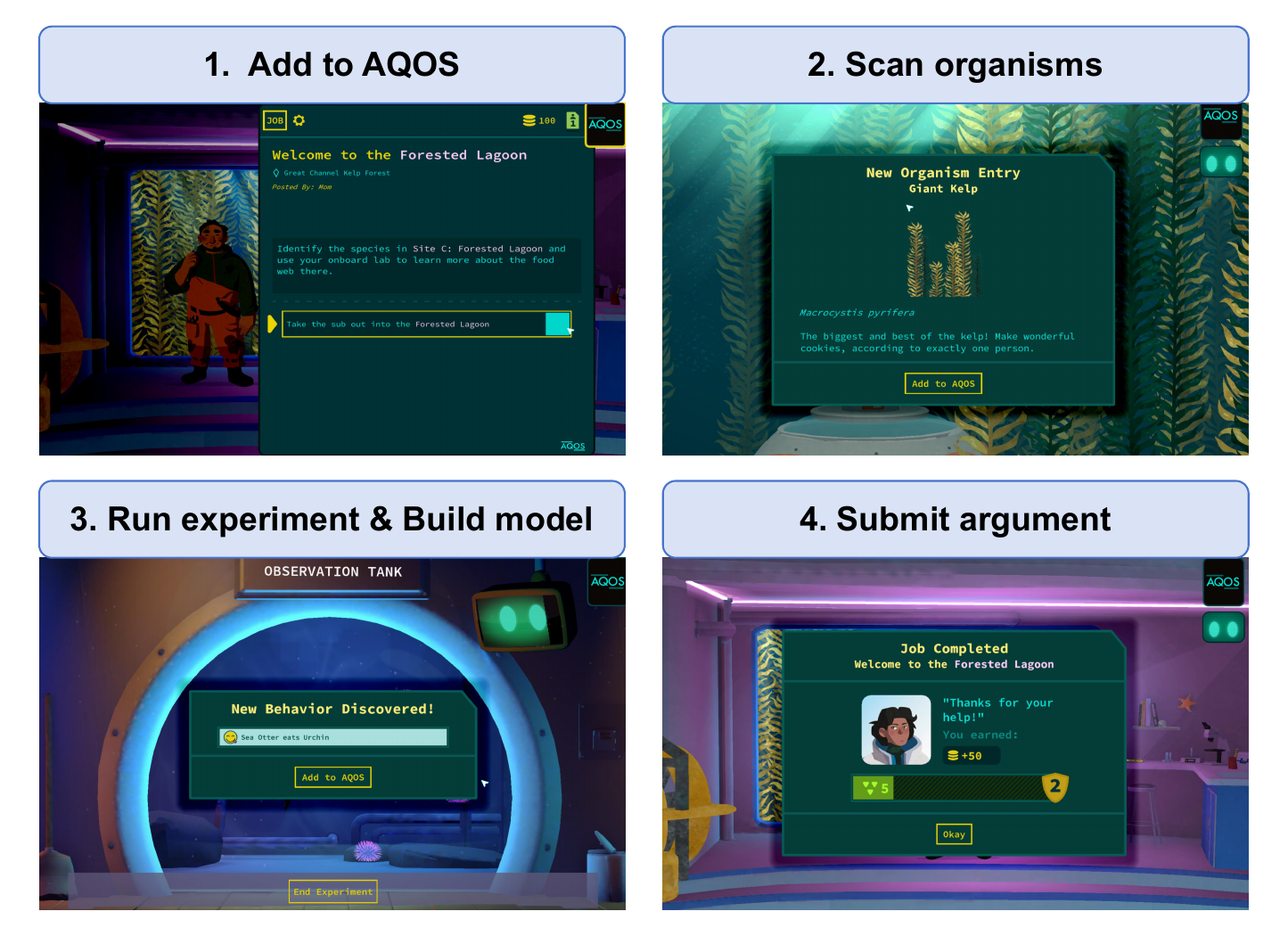}
  \caption{Interface of the AQUALAB educational science game}
  \label{fig:aqualab_interface}
\end{figure}

AQUALAB is a science-learning game in which players complete jobs and constituent tasks by collecting evidence, running experiments, building ecosystem models, and submitting scientific arguments. The game interface is shown in Figure~\ref{fig:aqualab_interface}.

The logged interactions therefore form a naturally sequential decision process: at each step, a player takes an in-game action, the game state is updated, and progress toward task completion may increase, stall, or reverse. We treat these logs as offline behavioral trajectories and convert them into standard transitions of the form $(s_t,a_t,r_t,s_{t+1})$, where the state representation summarizes the current scene, job and task status, experimental settings, accumulated observations, model state, and currently available tools.

Because the raw gameplay records contain many exploratory clicks and other low-information interactions, the empirical analysis uses a filtered episode-level dataset rather than the full log stream. We first construct cumulative-reward trajectories from the raw transition file and remove episodes that are too short or otherwise weakly informative. For long episodes, we restrict attention to the later portion of the trajectory, where actions are more directly tied to outcome-relevant progress. We then apply a step-level filter based on changes in cumulative reward, retaining transitions only when the absolute stepwise change satisfies $|\Delta_t| \ge 1.0$, while always preserving the initial and final steps of each episode to maintain trajectory integrity. This broad-to-narrow preprocessing pipeline is designed to suppress flat or incidental behavior and to retain decision steps that are more directly tied to measurable task progress.

The final analytic sample contains 1,133 episodes and 8,474 transitions, reduced from 106,315 raw transitions. We use the episode, rather than the player, as the primary analysis unit because some users appear in multiple gameplay sessions and the available logs do not always support a clean aggregation of all behavior into a single homogeneous trajectory. Estimation is based on the same block-coordinate MAP procedure introduced in Section~\ref{sec:method}, using the filtered AQUALAB trajectories to jointly learn the shared value function and the episode-level latent ability parameters $\hat{\beta}$.

\subsection{Association Between Estimated Ability and Cumulative Reward}
\label{subsec:aqualab_alignment}

To evaluate whether the estimated latent parameter has a plausible behavioral interpretation, we examine its association with cumulative trajectory reward. This analysis is intended as a validity check rather than as evidence that $\hat{\beta}$ represents a pure ability construct. In an open-ended environment such as AQUALAB, reward accumulation reflects not only effective decision making but also heterogeneous exploration styles, local detours, and the fact that some players may progress slowly but still engage productively with the task. Even so, if the RLMM is capturing a substantively meaningful performance-related signal, episodes with larger estimated $\hat{\beta}$ should on average achieve higher cumulative reward.

After 20 block-coordinate MAP iterations on the filtered dataset, the correlation between $\hat{\beta}$ and cumulative trajectory reward was 0.5842 by Pearson correlation and 0.5760 by Spearman correlation. The close agreement between the linear and rank-based summaries indicates that the positive association is not driven solely by a few extreme trajectories. At the same time, the association is far from deterministic, which is consistent with the exploratory nature of AQUALAB and with the fact that cumulative reward also reflects detours, heterogeneous strategies, and other sources of behavioral variation. Taken together, these results suggest that the estimated $\hat{\beta}$ captures an important performance-related component of behavior, without exhausting all of the meaningful variation in the observed trajectories. Table~\ref{tab:aqualab_alignment} summarizes these alignment statistics.

\begin{table}[t]
  \centering
  \caption{Behavioral alignment of the estimated ability parameter with cumulative trajectory reward in the AQUALAB data}
  \label{tab:aqualab_alignment}
  \begin{tabular}{lcc}
    \toprule
    Model & Pearson & Spearman \\
    \midrule
    RLMM & 0.5842 & 0.5760 \\
    \bottomrule
  \end{tabular}
\end{table}

We do not include a full MDP-MM baseline in this empirical study. In its original form, the MDP-MM assumes fixed task dynamics and reward structure, a fully enumerated tabular state space, and repeated dynamic-programming updates across candidate values of the person parameter. These assumptions were reasonable for the small peg-solitaire environments in Section~\ref{sec:experiments}, but they are difficult to justify and impractical to maintain for the large, noisy, and open-ended AQUALAB trajectories. The RLMM avoids that bottleneck by learning a shared parametric value function directly from logged experience, which makes estimation feasible on the filtered AQUALAB dataset while still producing person-level latent parameters with a meaningful positive association with observed behavioral performance.

\subsection{Completion and efficiency by estimated ability}
\label{subsec:aqualab_interpretation}

We next examine how $\hat{\beta}$ relates to two directly interpretable features of AQUALAB performance: completion and efficiency. These episode-level summaries are computed on the full raw episodes and then paired with the corresponding $\hat{\beta}$ estimates from the filtered-data fit. The analysis is descriptive, but it helps clarify what kinds of behavioral differences are associated with larger estimated person parameters in this environment.

\begin{figure}[tbp]
  \centering
  \includegraphics[width=0.65\linewidth]{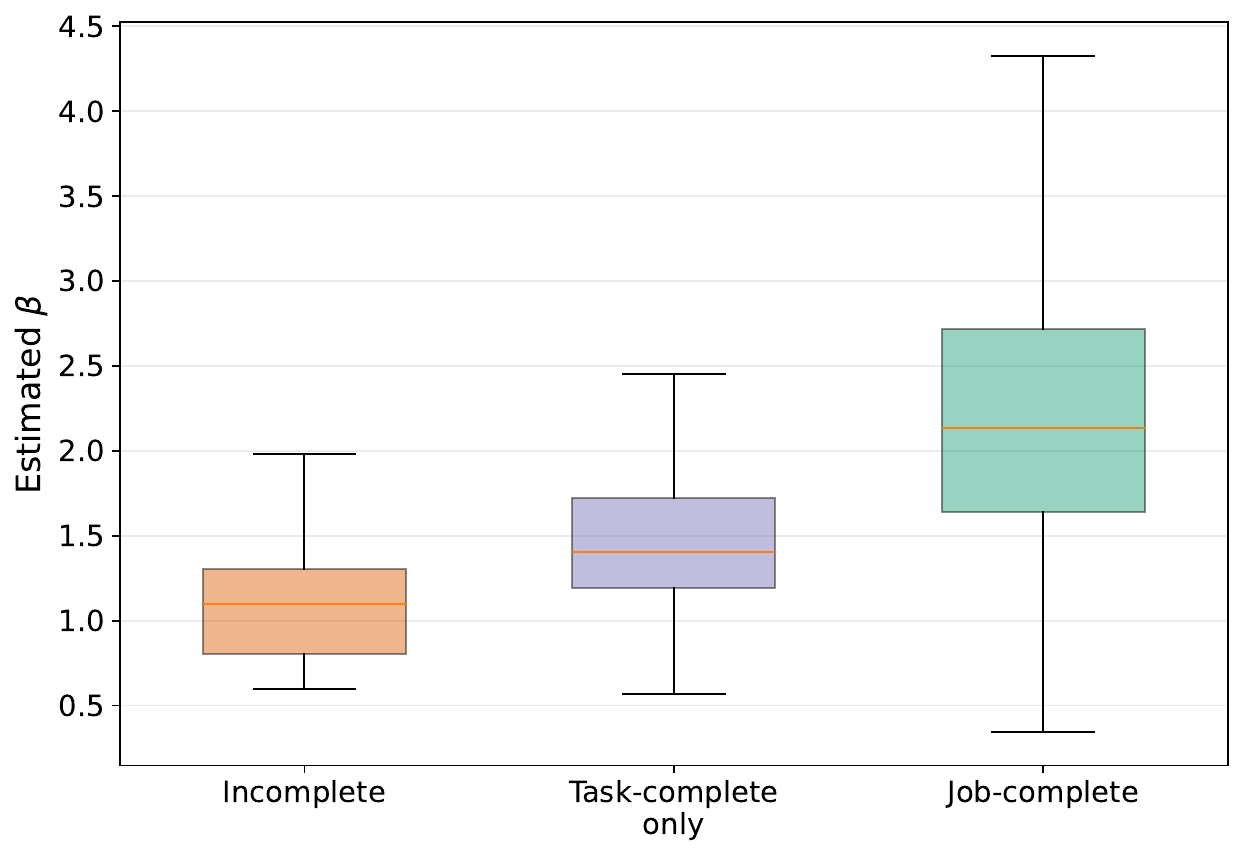}
  \caption{Distribution of the estimated ability parameter $\hat{\beta}$ by episode completion status in AQUALAB}
  \label{fig:aqualab_completion_boxplot}
\end{figure}

A first pattern appears in episode completion status. As shown in Figure~\ref{fig:aqualab_completion_boxplot}, the distribution of $\hat{\beta}$ is ordered across episodes that remained incomplete, episodes that completed some tasks but not the full job, and episodes that reached job completion. The corresponding mean estimates were 1.181, 1.506, and 2.343, respectively, and the overall group difference was highly significant under a Kruskal--Wallis test ($p = 8.62 \times 10^{-29}$). Episodes with larger estimated person parameters therefore tend to show more complete task execution.

\begin{figure}[tbp]
  \centering
  \includegraphics[width=0.95\linewidth]{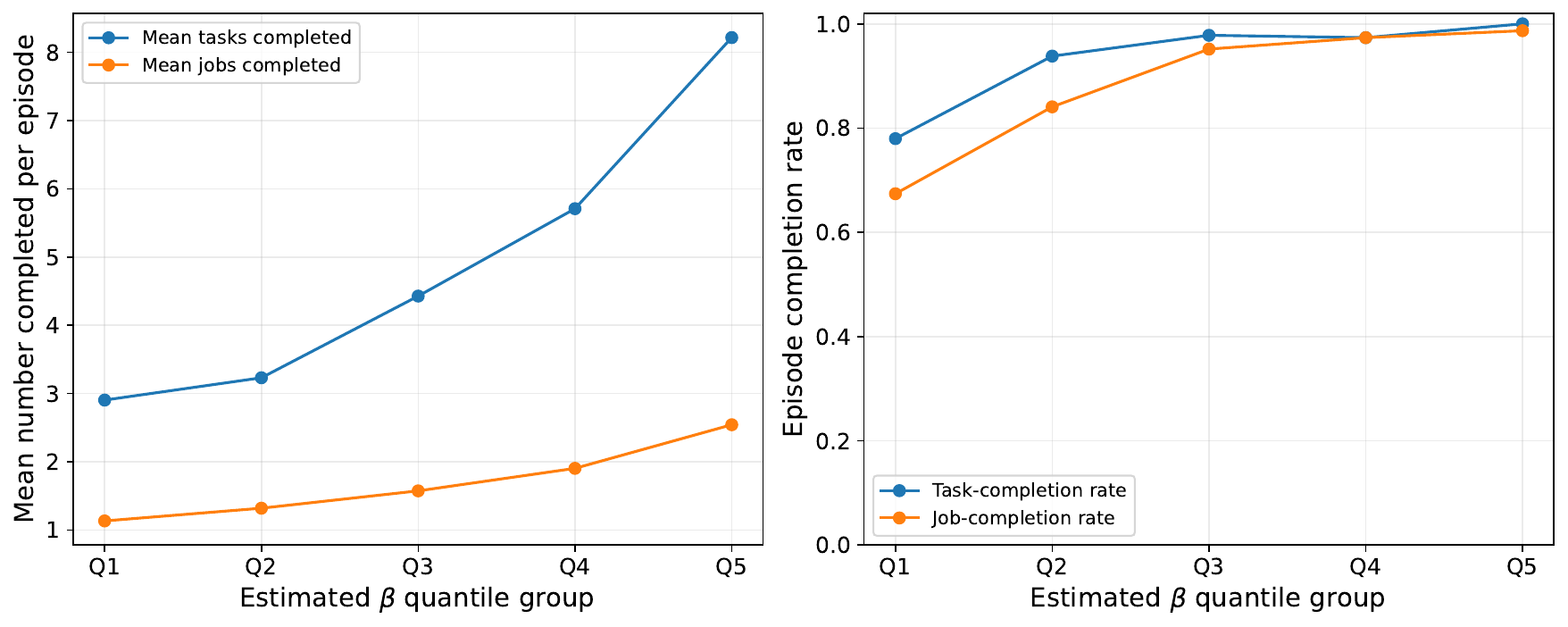}
  \caption{Completion summaries across quintiles of $\hat{\beta}$ in AQUALAB}
  \label{fig:aqualab_completion_trend}
\end{figure}

The same pattern appears when episodes are ordered more continuously by estimated ability. Figure~\ref{fig:aqualab_completion_trend} shows that, from the lowest to the highest $\hat{\beta}$ quintile, both the number of completed tasks and the number of completed jobs increase, and the corresponding completion rates rise as well. This suggests that the completion difference is not limited to a few separated groups, but extends across the broader $\hat{\beta}$ distribution.

\begin{figure}[tbp]
  \centering
  \includegraphics[width=0.65\linewidth]{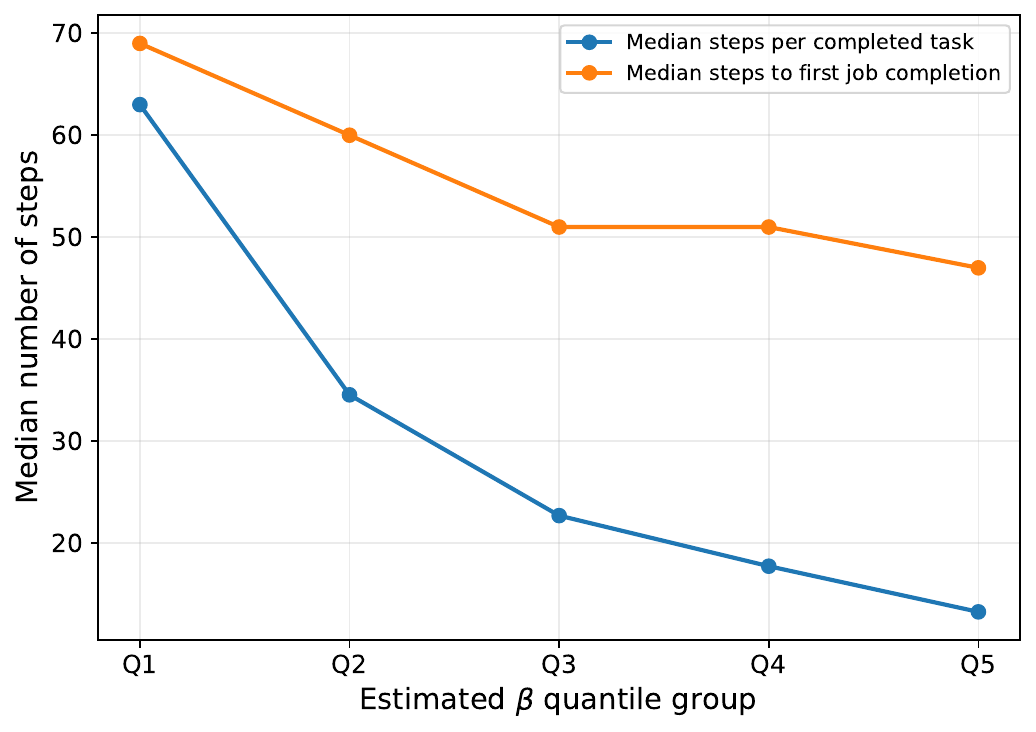}
  \caption{Efficiency summaries across quintiles of $\hat{\beta}$ in AQUALAB}
  \label{fig:aqualab_efficiency}
\end{figure}

Efficiency summaries point in the same direction. Figure~\ref{fig:aqualab_efficiency} shows that episodes in higher $\hat{\beta}$ quantiles generally require fewer interaction steps to achieve successful progress. The clearest pattern appears in steps per completed task, which declines steadily as $\hat{\beta}$ increases. Steps to first job completion also tend to decline across quantiles, although that pattern is naturally harder to interpret because it is defined only for episodes that eventually complete a job.

Overall, episodes with higher $\hat{\beta}$ tend to complete more of the game content and to do so more efficiently. These regularities help clarify part of the behavioral meaning of the estimated parameter, while leaving room for other sources of variation in this environment.

\section{Discussion}
\label{sec:discussion}

The present study proposed the Reinforcement Learning Measurement Model (RLMM) as a scalable extension of the Markov decision process measurement model \citep[MDP-MM,][]{lamar2018markov}. The main contribution of the RLMM is to reformulate the MDP-MM around a shared parametric action-value function while preserving the interpretation of the person parameter $\beta$ as value-based choice consistency. This reformulation addresses the main computational bottleneck of the original tabular approach without turning the model into a purely predictive model of sequential behavior. Instead, $\beta$ remains a psychometric parameter linked to how consistently a person selects actions with higher estimated value. The resulting framework combines a Boltzmann choice rule, a soft Bellman-consistency penalty, and block-coordinate MAP estimation. It also yields step-level influence diagnostics for locating decision points that matter most for person estimation.

The simulation results support this reformulation in both measurement and computational terms. Across the four peg-solitaire boards originally evaluated by \citet{lamar2018markov}, the RLMM recovered person parameters more accurately than the tabular MDP-MM baseline. It was also more computationally stable: the tabular baseline became markedly slower as task complexity increased, whereas the RLMM maintained relatively stable runtime across the same boards. The larger-board experiments sharpen this contrast by moving into a regime where repeated tabular dynamic programming is no longer practical. In those settings, the RLMM remained estimable and still produced reasonable recovery of the latent parameter. The empirical AQUALAB analysis cannot provide the same ground-truth recovery evidence, but the positive associations of $\hat{\beta}$ with cumulative reward, completion, and efficiency suggest that the estimated parameter captures a meaningful component of performance in a more open-ended environment.

In open-ended task environments, estimates of individual person parameters depend on how researchers define states, actions, and rewards within the model. These modeling choices are less transparent in unstructured environments than in constrained tasks such as peg solitaire. The current RLMM also uses a single positive ability parameter. This parameter is easy to interpret, but interactive behavior may vary along dimensions that a one-dimensional latent variable cannot fully capture, including strategy selection, task persistence, and exploration patterns.

These considerations point to several promising directions for future work. First, the RLMM can be extended to multidimensional or hierarchical formulations, particularly in settings where repeated task episodes from the same individual are available, to capture more nuanced differences in ability and within-person change over time. Second, the framework can be adapted to partially observed environments and richer, high-dimensional state representations, where function approximation is not just a practical convenience but a necessary methodological requirement. More broadly, the RLMM demonstrates that core reinforcement learning ideas can be integrated into psychometric models of process data while retaining a clear, interpretable measurement orientation. This balance between flexibility and measurement validity is challenging to maintain, but it is essential if sequential behavioral data are to support defensible, interpretable assessment rather than prediction alone.

\section*{Funding}
F.J. was supported by the Seed Grant of the International Network of Educational Institutes (Grant No. 522011), the Connaught Fund (Grant No. 520245), and the Social Sciences and Humanities Research Council of Canada (SSHRC; Grant No. 215119; Canada Research Chair: CRC-2024-00169).

\section*{Competing Interests}
The authors declare no competing interests.

\section*{Data Availability}
The AQUALAB gameplay logs analyzed in the empirical study were obtained from the February 2026 export of Wake: Tales from the Aqualab released through Open Game Data by Field Day Lab: \url{https://opengamedata.fielddaylab.wisc.edu/gamedata.php?game=AQUALAB}. The Open Game Data page provides access to raw time-sequenced gameplay data and derived data resources for the game.

\printbibliography

\end{document}